\documentclass{elsart}

\usepackage{psfig}

\usepackage{natbib}

\begin{document}

\runauthor{Cicero, Caesar and Vergil}


\begin{frontmatter}

\title{X-ray Continuum Slope and X-ray Spectral Features in NLS1 Galaxies}

\author{Hagai Netzer}
\address{School of Physics and Astronomy and the Wise Observatory,
Tel Aviv University, Tel Aviv 69978, Israel}

\begin{abstract}
The idea that some of the unusual features in the X-ray spectra of Narrow-Line
Seyfert 1 galaxies (NLS1s) are due to the  steep X-ray continuum 
is tested by comparing photoionization model calculations with various observed properties 
of Seyfert 1 galaxies. A meaningful comparison must involve the careful use of the ``right''
 X-ray ionization
parameter, designated here  $U(oxygen)$.
 When this is done, it is found that the strength of the continuum
absorption features is insensitive to the exact slope of the 0.1--50 keV continuum. It
is also shown that the complex of iron L-shell lines near 1 keV can produce 
strong absorption  and emission features, depending on the gas distribution and line widths.
While this may explain some
unusual X-ray features in AGN,  the predicted intensity of the features
do not distinguish NLS1s from broader line sources. Finally, acceleration of
highly ionized gas, by X-ray radiation pressure, is also not sensitive to the
exact slope of the X-ray continuum.
\end{abstract}

\begin{keyword}
galaxies: active; quasars: general; quasars: absorption lines; X-rays: galaxies
\end{keyword}

\end{frontmatter}


\section{Introduction}
The steep X-ray spectrum of Narrow Line Seyfert 1 galaxies (NLS1s) has been shown
to be highly correlated with many of the unusual properties of these sources
(e.g. Boller, Leighly and Wills' articles in this volume). It is therefore important to study
the consequences of this continuum shape and its influence on the highly ionized gas (hereafter
HIG) in such sources. In particular, it is interesting to test
the idea that steep slope active galactic nuclei (AGN)
  contain HIG which is either significantly
more ionized or significantly more neutral than the same component in broad line Seyfert 1 galaxies
(BLS1s). If correct, this would have significant influence on the observed
  X-ray features and,
perhaps, also on the properties of the associated UV absorption lines.

This paper presents 
 the results of new model calculations pertaining to the strength
(i.e. the optical depth) of the continuum absorption features around 0.7-0.9 keV, the
absorption and emission lines in the two sub-classes of Seyfert 1 galaxies, 
and the motion of the HIG in NLS1s. In what follows, the 
dividing line between NLS1s and BLS1s is defined at 
 X-ray continuum photon slope
of $\Gamma=2.3$.

\section{Unusual spectral features  in NLS1s}
\subsection{A comparison of the X-ray spectrum of NLS1s and BLS1s}
A large number of BLS1s have been observed by ASCA, allowing a detailed investigation
of the X-ray absorption features and  some statistical analysis of 
 these properties (e.g. Reynolds 1997; George et al. 1998). 
This, however, is not the case for NLS1s. The
objects studied so far from this group are few and the signal-to-noise of most
 of the ASCA spectra is far
inferior to the high quality spectra of  the brightest BLS1s. The information available
in the literature, as well as the new data presented in this meeting, allow however,
a superficial comparison of the X-ray spectral properties of the two groups. 
In particular, it was claimed that:
\begin{itemize}
\item
Some NLS1 galaxies show a strong absorption feature at around 1 keV 
which is different in shape and
in energy from the commonly observed O~VII and O~VIII continuum
absorption features in BLS1s. This
 was interpreted
as due to  O~VII and O~VIII resonance  absorption lines in a gas 
moving at a relativistic speed away from
the central object (Leighly et al 1997). 
\item
Other NLS1s (e.g. Akn~564, see Turner, Netzer \& George 1999) show strong emission near 1 keV
and no sign of X-ray absorption over the ASCA energy range. 
The strength and energy of
this emission feature is still a source of discussion. Turner et al.
(1999) have suggested that it may be produced by a large number of iron
L-shell lines indicating, perhaps,  iron over-abundance.
\item
 The 1 keV absorption feature observed  in some
 NLS1s is the result of a large number of iron absorption
lines close to this energy (Nicastro, Fiore \& Matt 1999). The strength of this feature and
the relative weakness of the bound-free to O~VII and O~VIII absorption, are
related to the unusually steep continuum in NLS1s.
\end{itemize}
The following is a closer examination of these claims. The underlying assumption
is that 
photoionization by the X-ray continuum is the sole excitation and heating source of the HIG in both classes of AGN.

\subsection{X-ray continuum absorption in NLS1 galaxies}

The idea that a steeper X-ray continuum results in a different level of ionization of
the surrounding gas can be tested  by photoionization models. However,
we must make sure that the comparison is meaningful and the  calculations
reflect, indeed, the influence of the X-ray continuum. In particular, it is important to
use the ``correct'' ionization parameter,
\begin{equation}
U=\int^{E_2}_{E_1} \frac{(L_E/E)dE}{4 \pi r^2 n_H c}
\end{equation}
i.e. to carefully choose the most appropriate energy range $E_1-E_2$.

Several different ionization parameters are currently in use, e.g. the UV ionization parameter designated here  $U(hydrogen)$ ($E_1=13.6$ eV), and the X-ray ionization
parameter (Netzer 1996) with $E_1=0.1$ keV and $E_2=10$ keV. 
The fractional ionization of O~VII and O~VIII, the ions contributing most to the
bound-free absorption by the HIG,  are determined,
almost exclusively, by $E>0.5$ keV photons. Hence, it is useful to define a new
ionization parameter, $U(oxygen)$, over the energy range corresponding to oxygen ionization,
 $E_1=0.538$ keV and $E_2=10$ keV. 
Extensive tests show that HIG clouds, similar in their properties to those observed in BLS1s, are hardly
affected by X-ray photons with energy below $E_1$. Hence, a meaningful comparison
of the effect of the continuum slope is through comparing models with the same $U(oxygen)$.
It can also be shown that some combinations of different $U(hydrogen)$ and 
spectral energy distribution (SED) can produce conflicting results regarding
the influence of the X-ray continuum slope.

Figure 1 shows a comparison of the spectra of two HIG clouds that are exposed
to  {\bf (a).} a typical BLS1 continuum
with $\Gamma=2$ and {\bf (b).} an extremely steep X-ray continuum of $\Gamma=2.8$. 
  $U(oxygen)$ is
 the same in both cases (0.02). This is about the average 
 value measured by George et al. (1998) for their sample of BLS1s.
 The softer part of this continuum is a combination of a power-law IR continuum
and a weak UV bump. This corresponds, for the case of $\Gamma=2.8$,
  to $U(hydrogen)=28$
and $U_X=0.4$.
The column density is typical of strong absorption HIG (10$^{22}$ cm$^{-2}$),
 the hydrogen number density, $n_h$,
is 10$^8$ cm$^{-3}$ (the model is insensitive to this parameter provided it is below about
10$^{13}$ cm$^{-3}$) and the composition close to solar. As evident from the 
diagram, ``standard
slope'' and ``steep slope'' continua produce roughly the same strength absorption features 
when normalized to the same $U(oxygen)$.
Thus, the X-ray continuum slope by itself is 
not the cause of the apparent difference in continuum
absorption properties between NLS1s and BLS1s.

\begin{figure}[htb]
\centerline{\psfig{figure=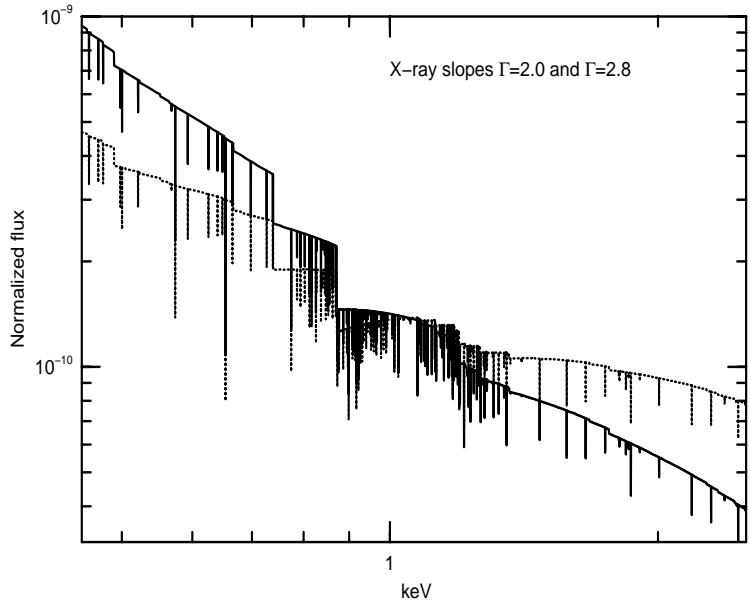,height=3.9truein,width=4.5truein,angle=0}}
\caption{X-ray absorption features for $U(oxygen)=0.02$ and two slopes, $\Gamma=2.0$ 
  (dotted line) and $\Gamma=2.8$ (solid line). Note the similarity of the strongest
  X-ray absorption features near 0.8--0.9 keV}
\end{figure}

\subsection{X-ray absorption lines in NLS1s}

Next we test the idea that the mysterious 1 keV absorption feature is due to the
conglomeration of a large number of iron L-shell lines combined with an unusually
steep continuum. The absorption spectrum of such gas has been calculated allowing for
various slope continua and various width lines (e.g. various values of the micro-turbulent velocity).
 The results indicate that  the optical depths and equivalent widths (EWs) of the
absorption lines are very insensitive to the continuum slope, when normalized to the
same value of $U(oxygen)$. Both BLS1 continua and NLS1 continua produce strong absorption lines
over the range of interest. The observed EWs depend on the line widths, the covering factor
and the turbulent velocity. For more discussion see Netzer (2000).
 
The examples shown in Fig. 2 and 3 are for a HIG illuminated by a $\Gamma=2.5$ X-ray continuum
with $U(oxygen)=0.2$ ($U(hydrogen)=143$ and $U_X=2.5$).  Fig. 2  is a pure absorption case, i.e. the $4 \pi$
covering factor
is very small but the line-of-sight covering factor is 1.0.
The softer part of the continuum, and the other model parameters,
are  identical to the ones used for
the previous case.
  As seen in the diagram, 
strong absorption features are indeed present. This confirms the Nicastro
et al. (1999) suggestion about the origin of the 1 keV absorption feature,
especially for gas clouds with large micro-turbulent velocity
(200 km/sec in the case shown here). Pure thermal profiles do not produce
strong enough absorption lines to explain the 1 keV feature reported
by Leighly et al. (1997). 

We note that the present
calculations are rather different from the ones presented by Nicastro et al. (1999)
for the same parameters. This is true for the H-like and He-like lines as well
as the the iron L-shell lines (Netzer 2000).
 As for the
influence of the X-ray slope, a comparison of various slope continua, with the
same value of $U(oxygen)$,  clearly show that this behavior
is typical of both NLS1s and BLS1s. Thus the strong absorption lines and
large absorption EWs are typical of the two groups of sources.

\begin{figure}[htb]
\centerline{\psfig{figure=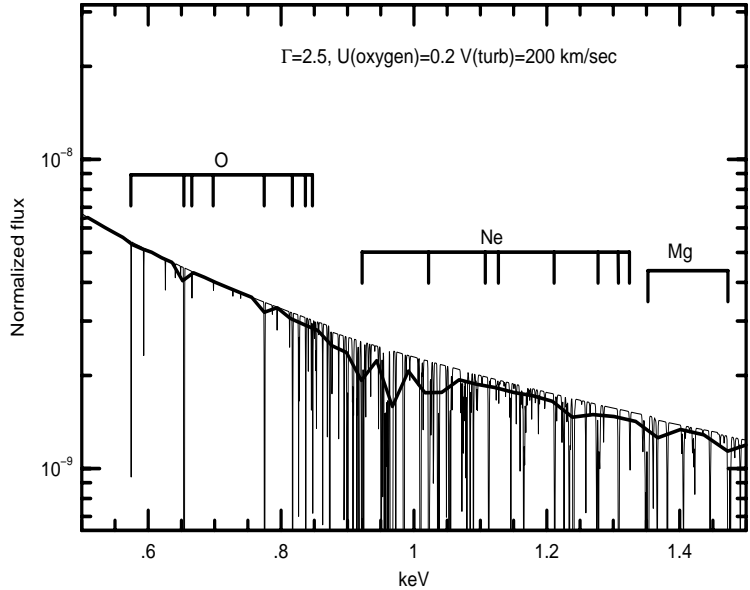,height=3.9truein,width=4.5truein,angle=0}}
\caption{X-ray absorption features for $U(oxygen)$=0.2  $\Gamma=2.5$ and different
spectral resolution: $E/\Delta E=1000$ and $E/\Delta E=20$. The low resolution is
similar to what is provided by CCD type detectors  at around 2 keV. It produces strong
and broad absorption features around 1 keV that can be confused with blueshifted
  components.}
\end{figure}

\subsection{X-ray emission lines in NLS1s}

Finally, the idea of an unusually strong emission feature near 1 keV in NLS1 spectra,
was also tested. Typical
X-ray lines in photoionized gas are weak with small  EWs (1-10 eV for the strongest lines
assuming typical HIG, see Netzer 1996). This
is well below the EW observed by Turner et al. (1999) in the ASCA spectrum of
Akn~564 (about 70 eV). However, the inner geometry of the source may be such that the $4 \pi$ covering
factor is large yet the line-of-sight is relatively clear. In this case, continuum absorption
is negligible and the scattered continuum photons will be seen on top of 
the unabsorbed
powerlaw continuum. Such an unusual geometry can appear, for example,
 in flat HIG systems with extreme inclination to the line-of-sight.

Fig. 3 shows the theoretical spectrum resulting from such a
 special geometry. The cloud
is identical to the one shown in Fig. 2 but the $4 \pi$ covering
factor is large (0.8) and the line-of-sight is clear of absorbing material. 
The emission around 0.9-1.0 keV is,
indeed, very strong. This is especially noticeable when plotted with the low ASCA resolution
(solid line). Note again that the turbulent velocity is the key factor and clouds with
pure thermal profiles  produce much weaker emission lines.

\begin{figure}[htb]
\centerline{\psfig{figure=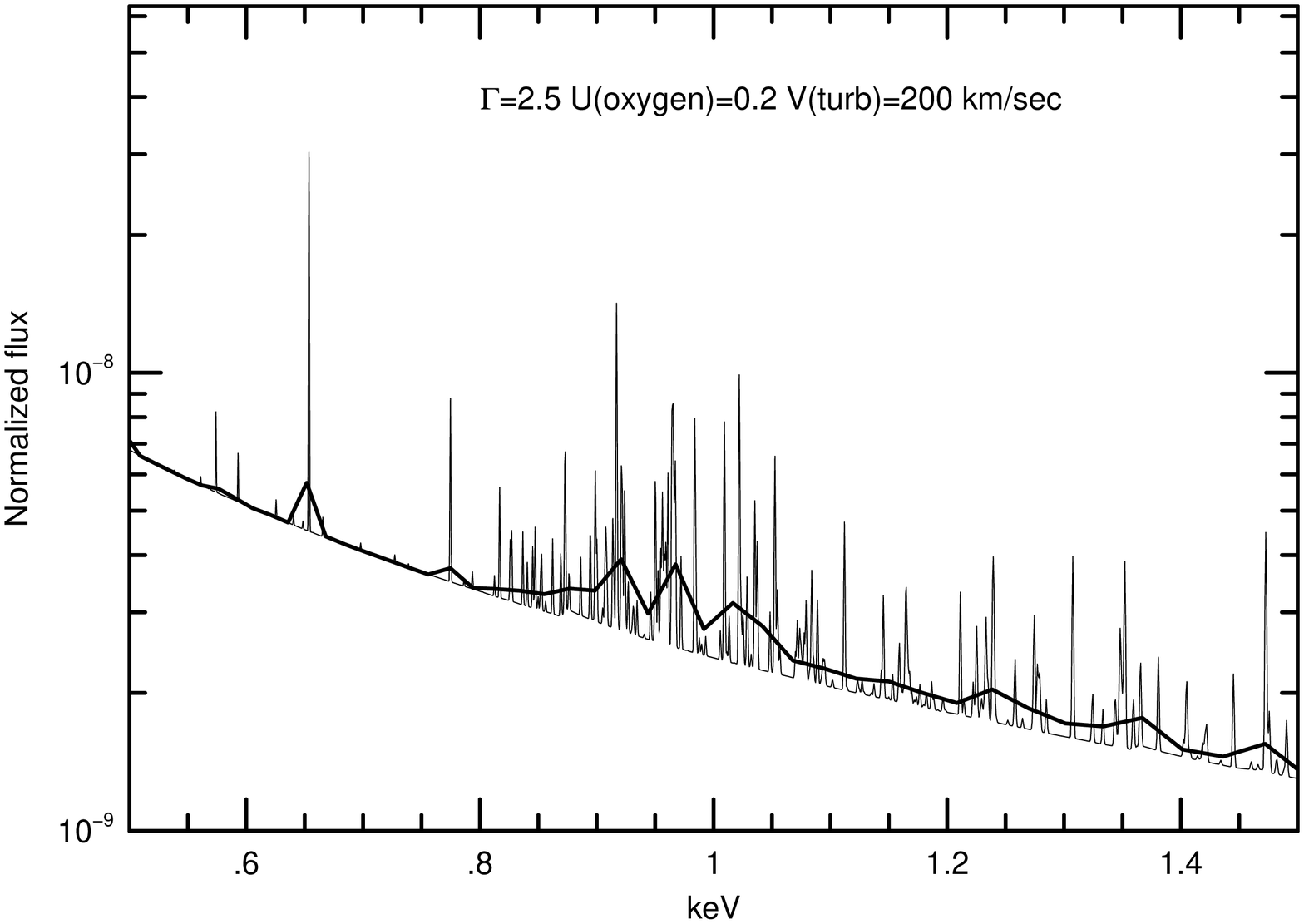,height=3.9truein,width=4.5truein,angle=0}}
\caption{The same gas as in Fig. 2 shown for the case of no absorption along the line of
sight. The total equivalent width of the strong and broad emission features near 0.9 keV,
  relative to the incident continuum, is about 60 eV}
\end{figure}

\section{Motion of the ionized gas}
The ionized nuclear gas in  Seyfert galaxies is subjected to the intense radiation field
of the central source 
which can produce strong radiation pressure forces. Such forces,  due
to the UV continuum, and their influence on the ionized gas dynamics,
 have been studied in detail for BALQSOs (e.g. Arav, Lee
\& Begelman, 1994 and references therein). Yet, little has been done so far
on the acceleration
of the HIG by the intense X-ray source.

In a recent paper, Chelouche and Netzer (2000) investigated the physics and 
dynamics of HIG clouds exposed to typical AGN X-ray continua. The results confirm that
such gas can be accelerated to high velocities depending on the
 origin of the flow relative to the center,
its column density, the confining pressure and the absorption line widths. Typical velocities of 500-1000
  km/sec have been obtained for gas clouds that originate just outside  the 
broad line region (BLR); a likely location of HIG clouds.

The  Chelouche and Netzer (2000) calculations have also been applied to steep spectrum sources,
in an attempt to check whether the gas dynamics can provide another
 distinguishing factor between
BLS1s and NLS1s. The detailed calculations, that will be presented elsewhere, show that
the less luminous steeper X-ray continua of NLS1s are as efficient   
as the shallower and more luminous BLS1 continua in accelerating the HIG to high velocities.
 Thus, the dynamics
of the HIG is not likely to provide a clear distinction between NLS1s and BLS1s.






\begin{thebibliography}{999}

\bibitem{A1} N. Arav, Z. Li, \& M. Begelman, {\em ApJ} {\bf 432} (1994) 62.

\bibitem{C1} D. Chelouche, \& H. Netzer {\em ApJ} {\bf in press} (2000).

\bibitem{M1} I.M. George, T.J. Turner, H. Netzer, K. Nandra, R.F. Mushotzky, \& T. Yaqoob,
   {\em ApJS} {\bf 114} (1998) 73.

\bibitem{1} K.M. Leighly, R.F. Mushotzky, K. Nandra, \& K., Forster, {\em ApJLett} {\bf 489} (1997) L25.

\bibitem{2} F. Nicastro, F. Fiore, \& G. Matt, {\em ApJ} {\bf 517} (1999) 108.
 

\bibitem{3} H. Netzer, {\em ApJ} {\bf 473} (1996) 781.

\bibitem{4} H. Netzer,  (2000) (in preparation).

\bibitem{5} C.S. Reynolds {\em MNRAS} {\bf 228} (1997) 513.

\bibitem{6} T.J. Turner, H. Netzer, \& I.M. George, {\em ApJ} {\bf 526} (1999) 52. 


\end{thebibliography}
\end{document}